\documentclass[aps,pra,reprint,longbibliography,superscriptaddress,nofootinbib]{revtex4-2}
\usepackage[T1]{fontenc}
\usepackage{amsmath,amssymb,graphicx}
\usepackage[colorlinks=true,citecolor=blue,linkcolor=blue,urlcolor=blue]{hyperref}
\hypersetup{pdftitle={Coherent heat-bath configuration interaction with many-body perturbation theory for multivalent atoms},pdfauthor={Shi-Cheng Yu, Zhen-Xiang Zhong, Cheng-Bin Li}}
\begin{document}
\title{Coherent heat-bath configuration interaction\texorpdfstring{\\}{ }with many-body perturbation theory for multivalent atoms}
\author{Shi-Cheng Yu}
\affiliation{Center for Theoretical Physics, School of Physics and Optoelectronic Engineering, Hainan University, Haikou 570228, China}
\author{Zhen-Xiang Zhong}
\email{zxzhong@hainanu.edu.cn}
\affiliation{Center for Theoretical Physics, School of Physics and Optoelectronic Engineering, Hainan University, Haikou 570228, China}
\author{Cheng-Bin Li}
\email{cbli@apm.ac.cn}
\affiliation{Innovation Academy for Precision Measurement Science and Technology, Chinese Academy of Sciences, Wuhan 430071, China}
\begin{abstract}
Combinatorial growth of configuration-interaction spaces limits accurate calculations as the valence-electron number and orbital space increase.
To reduce the size of the configuration-interaction space, we develop coherent heat-bath configuration interaction in a configuration-state-function basis and combine it with many-body perturbation theory.
Configurations are selected by thresholding their coherent couplings to target wave functions, and a perturbative correction recovers residual valence correlation.
For Al, a stringent threshold gives about 16--37-fold compression with excitation-energy deviations from the uncompressed calculations below $0.74\,\mathrm{cm}^{-1}$; a looser threshold gives 63--113-fold compression with a maximum deviation of $18.22\,\mathrm{cm}^{-1}$.
For Si and P, roughly threefold to eightfold compression is achieved with maximum relative deviations of approximately 0.70\% and 0.32\%, respectively, for optical excitation energies.
The perturbative correction improves the accuracy of all calculated total energies.
These results support extending the method to high-precision calculations in more complex atomic systems.
\end{abstract}
\maketitle
\section{Introduction}
\label{sec:introduction}

Accurate atomic structure calculations support the extraction of nuclear properties from spectroscopy and the development of optical frequency standards~\cite{Lu2013NuclearRMP,Berengut2025IsotopeReview,Mitroy2010Polarizabilities,Ludlow2015Clocks}.
Electronic factors connect hyperfine measurements to nuclear moments~\cite{Yang2023LaserNuclei,Porsev2021ThMoments,Zitzer2025ThSpectroscopy} and isotope shifts to changes in nuclear charge radii~\cite{Skripnikov2024AlRadius,Lu2024LuRadius}.
For optical clocks, calculated atomic properties guide systematic-shift evaluations~\cite{Kozlov2018HCIClocks} and the search for new candidates~\cite{Yu2024P3Clocks,Dzuba2024CmBk}, while the polarizability-scale method~\cite{Barrett2019Polarizability,Wei2024Polarizability} and tune-out wavelengths~\cite{Safronova2015Tuneout,Heinz2020SrLattice} help characterize and control light shifts in existing clocks.
These applications require an accurate treatment of electron correlation.

Configuration interaction (CI) describes correlation beyond the mean-field approximation by expanding an atomic state in many-electron configurations~\cite{Eriksen2021FCIReview}.
Including all configurations gives the full-CI solution for the chosen Hamiltonian.
The expansion grows combinatorially, with approximately exponential scaling as the number of correlated electrons and the orbital space increase.
Selected CI combined with perturbation theory reduces this cost by retaining the configurations most important to the target states and recovering the remaining energy contribution perturbatively~\cite{Holmes2016HCI,Sharma2017SHCI}.
The central choice is how to measure configuration importance and construct a space that describes the required levels.

Atomic calculations have addressed this choice through several complementary approaches.
Neural networks select configuration state functions (CSFs) or relativistic configurations from approximate wave-function weights~\cite{Bilous2023GRASP,Bilous2024AtomicNN}, while configuration state function generators organize and select groups sharing angular-coupling structure~\cite{Li2023CSFG,Si2025GRASPG,Shi2026CSFGNN}.
Configuration interaction with perturbation theory (CIPT) includes external configurations through an effective interaction~\cite{Dzuba2017CIPT}.
The subsequent emu CI approach, used with CI and many-body perturbation theory (CI--MBPT), retains their couplings to an important subspace in the diagonalization and treats those retained couplings nonperturbatively~\cite{Geddes2018EmuCI,Kahl2019AMBiT}.

Perturbative selected-CI methods offer an adaptive way to construct the important subspace, using Hamiltonian couplings and the evolving target wave functions to guide configuration selection.
In configuration interaction by perturbatively selecting iteratively (CIPSI), importance is estimated from the first-order amplitude $\left|\sum_i H_{ai}c_i/(E-H_{aa})\right|$, whereas heat-bath configuration interaction (HCI) uses the simpler measure $\max_i|H_{ai}c_i|$~\cite{Holmes2016HCI}.
Here, $c_i$ is a coefficient in the current variational wave function, $E$ is its energy, and $H_{ai}$ and $H_{aa}$ are Hamiltonian matrix elements involving external determinant $a$.
The coherent sum $\sum_i H_{ai}c_i$ is thus already present in perturbative selection, as well as in the Epstein--Nesbet (EN) second-order energy correction used after HCI selection~\cite{Holmes2016HCI,Sharma2017SHCI}.
The framework also extends to excited states and selection for several target levels~\cite{Holmes2017ExcitedHCI}.
Applying an energy threshold directly to this full numerator gives a selection measure that retains interference between contributions from the current expansion.

Selected CI has also been extended to relativistic Hamiltonians and symmetry-adapted bases.
Wang and Sharma extended HCI to two- and four-component relativistic Hamiltonians in a spinor-determinant basis~\cite{Wang2023RelativisticSHCI}.
Prentice et al. implemented several selection schemes in both determinant and spin-adapted CSF bases, including state-averaged calculations~\cite{Prentice2023ModularSCI}.
For atomic structure, these developments motivate combining coupling-based selection with CSFs of definite total angular momentum and parity $J^\pi$, selection of whole relativistic configurations, and an effective Hamiltonian incorporating core correlation.

Here we formulate coherent heat-bath configuration interaction (CHCI) within this atomic framework and combine it with MBPT.
The method uses coherent CSF couplings to select whole relativistic configurations for one or several target states and recovers the remaining valence correlation through an EN correction.
We use the three-valence-electron full-CI space of Al to examine threshold scaling, compression, and iterative convergence, then extend the tests to the two-reference single- and double-excitation spaces of Si and P.
These comparisons assess the accuracy of the selected expansions and examine how correlation structure and the sharing of configurations among target states affect compression.
\section{Methods}
\label{sec:methods}

We have developed a CI--MBPT program in Julia and incorporated CHCI for configuration selection.
The methods implemented in the program are described below.
Hartree atomic units are used throughout this section.

\subsection{CI--MBPT}
\label{sec:ci-mbpt}

CI--MBPT combines a CI treatment of valence correlation with a perturbative treatment of core excitations~\cite{Dzuba1996CIMBPT,Kahl2019AMBiT}.
The method partitions the atom into a closed-shell core and $N_v$ active valence electrons.
The occupied core orbitals are obtained from a self-consistent Dirac--Hartree--Fock (DHF) calculation.
The resulting one-electron operator is
\begin{equation}
 h_{\mathrm{DHF}}=c\boldsymbol{\alpha}\cdot\mathbf p
 +(\beta-1)c^2+V_{\mathrm{nuc}}+V_{\mathrm{core}},
 \label{eq:dhf}
\end{equation}
where $c$ is the speed of light, $\boldsymbol{\alpha}$ and $\beta$ are Dirac matrices, $\mathbf p$ is the momentum operator, $V_{\mathrm{nuc}}$ is the nuclear potential, and $V_{\mathrm{core}}$ is the self-consistent core potential.
Valence and virtual orbitals are constructed by diagonalizing this operator in a B-spline basis in the same core potential~\cite{Kahl2019AMBiT}.
These orbitals provide the one-electron basis for both the valence CI expansion and the sums over virtual excitations of the core.

Virtual core excitations are incorporated through second-order MBPT, whose Goldstone diagrams represent self-energy corrections to the valence Hamiltonian~\cite{Dzuba1996CIMBPT,Kahl2019AMBiT}.
In these diagrams, the core is excited in an intermediate state and returns to its closed-shell configuration, leaving an effective interaction among the valence electrons.
Diagrams with one external valence electron generate the one-electron component $\Sigma_1$ of the self-energy operator, while those with two generate its two-electron component $\Sigma_2$, which includes screening by the core.
For atoms with three or more valence electrons, second-order diagrams also produce a three-electron component $\Sigma_3$~\cite{Berengut2011Chromium}.
We evaluate the one- and two-body corrections using the expressions given in Appendix B of Ref.~\cite{Berengut2006Carbon}.
The valence energies entering the MBPT denominators are fixed at the lowest valence orbital energy for each relativistic angular symmetry.

The resulting effective Hamiltonian for the valence electrons is
\begin{equation}
 H_{\mathrm{eff}}=\sum_{p=1}^{N_v}h_{\mathrm{DHF}}(p)
 +\sum_{p<q}g(p,q)+\sum_{n=1}^{3}\Sigma_n,
\label{eq:heff}
\end{equation}
where $p$ and $q$ label valence electrons, and $\Sigma_n$ denotes the full induced $n$-electron operator, including the sum over the corresponding valence electrons.
The interaction $g(p,q)=r_{pq}^{-1}+B_{pq}$ contains the Coulomb term, with $r_{pq}$ the distance between electrons $p$ and $q$, and the frequency-independent Breit operator $B_{pq}$.
The core mean field and residual interactions include the Breit contribution consistently.
For $\Sigma_3$, we retain matrix elements for which at least one of the two configurations belongs to the reference set.
The same effective Hamiltonian is used in the CI--MBPT and CHCI--MBPT calculations.

Slater determinants are constructed from the relativistic one-electron orbitals.
Linear combinations of these determinants with definite total angular momentum $J$ and parity $\pi$ form the orthonormal CSFs $|\Phi_i(J^\pi)\rangle$, where $i$ labels the CSFs~\cite{Kahl2019AMBiT}.
The parent CI space $\mathcal C$ comprises the CSFs generated from the reference configurations and the allowed excitations.
Within a given $J^\pi$ block, the wave function of atomic level $k$ is expanded as
\begin{equation}
 |\Psi_k\rangle=\sum_{i\in\mathcal C}C_{ik}|\Phi_i(J^\pi)\rangle,
 \label{eq:ci-expansion}
\end{equation}
where $C_{ik}$ is the expansion coefficient of CSF $i$ in that level.
The coefficients and the corresponding energy $E_k$ are obtained by solving the CI equation
\begin{equation}
 \sum_{j\in\mathcal C}H_{ij}C_{jk}=E_kC_{ik},
\label{eq:ci}
\end{equation}
where $H_{ij}=\langle\Phi_i|H_{\mathrm{eff}}|\Phi_j\rangle$ is the effective Hamiltonian matrix element between CSFs $i$ and $j$.
Each relativistic configuration $G$ specifies the occupations of the one-electron orbitals and can generate several CSFs within the same $J^\pi$ block.
In the following, $a\in G$ means that CSF $a$ belongs to configuration $G$.
CHCI selects configurations from $\mathcal C$ to construct a smaller space for diagonalization.

\begin{figure}[t]
\centering
\includegraphics[width=\columnwidth]{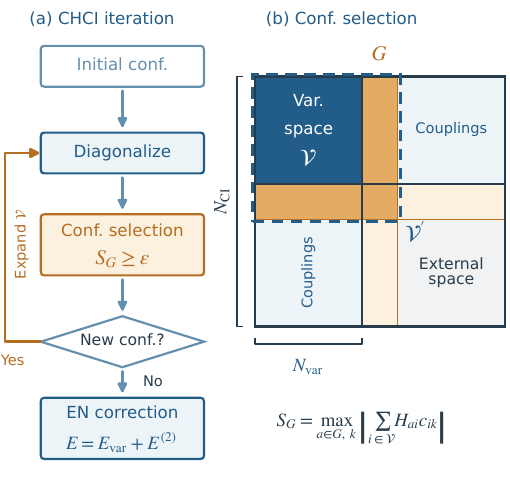}
\caption{Schematic illustration of CHCI. (a) Selection and diagonalization
repeat until no new configurations are added, followed by the external-space
Epstein--Nesbet (EN) energy correction. (b) Matrix illustration of the configuration
selection step in (a). Selecting configuration $G$ adds all its configuration state functions (CSFs) to the variational space $\mathcal V$,
forming $\mathcal V'$ (dashed outline). $N_{\mathrm{CI}}$ and $N_{\mathrm{var}}$
are the parent and variational CSF-space dimensions. $S_G$ is the largest
coherent coupling magnitude over CSFs $a\in G$ and target states $k$.}
\label{fig:chci}
\end{figure}

\subsection{CHCI--MBPT}
\label{sec:chci}

Let $\mathcal V\subseteq\mathcal C$ be the current variational space.
For target level $k$, diagonalization in $\mathcal V$ gives the variational wave function
\begin{equation}
 |\Psi_k^{\mathrm{var}}\rangle
   =\sum_{i\in\mathcal V}c_{ik}|\Phi_i\rangle,
 \label{eq:variational-state}
\end{equation}
where $c_{ik}$ are its expansion coefficients in the selected space, normalized by $\sum_{i\in\mathcal V}|c_{ik}|^2=1$.
The variational energy $E_k^{\mathrm{var}}$ is the eigenvalue of $H_{\mathrm{eff}}$ restricted to $\mathcal V$ for this target state.
For an external CSF $a\in\mathcal C\setminus\mathcal V$, we define its coupling to target state $k$ as
\begin{equation}
 \sigma_{ak}
 =\langle\Phi_a|H_{\mathrm{eff}}|\Psi_k^{\mathrm{var}}\rangle
 =\sum_{i\in\mathcal V}H_{ai}c_{ik}.
 \label{eq:coherent-coupling}
\end{equation}
The contributions from all components of the current wave function are summed before taking the absolute value.
Their relative signs are therefore retained, including constructive and destructive interference in the coupling to an external CSF.
This coherent amplitude determines both configuration selection and the external-space energy correction.
It is the numerator of the first-order perturbative amplitude used in perturbative selected CI~\cite{Holmes2016HCI}.
Here, the selection threshold is applied directly to its magnitude, with the configuration-level and multistate rule defined below.

CHCI uses a relativistic configuration as the elementary unit of selection.
CSFs belonging to the same configuration share the same orbital occupations and differ in their angular coupling, so we treat them as physically related components of the atomic state.
Selecting whole configurations also fits the architecture of our program, in which the CI matrix is generated from a configuration list.
We therefore assign an importance measure $S_G$ to each external configuration $G$.
For a set of target levels $\mathcal K$ in the same $J^\pi$ block, this measure is
\begin{equation}
 S_G=\max_{\substack{a\in G\\k\in\mathcal K}}|\sigma_{ak}|.
 \label{eq:configuration-selection}
\end{equation}
The maximum is taken jointly over the CSFs belonging to $G$ and the target states.
Given an energy threshold $\epsilon$, a configuration is selected when $S_G\geq\epsilon$, so one pair $(a,k)$ reaching the threshold is sufficient.
All CSFs of that configuration within the symmetry block are then added to $\mathcal V$.
Figure~\ref{fig:chci}(b) illustrates this step through the orange rows and columns associated with $G$ and the dashed boundary of the enlarged space $\mathcal V'$.
The dimensions $N_{\mathrm{CI}}$ and $N_{\mathrm{var}}$ count CSFs in $\mathcal C$ and $\mathcal V$, respectively.
Since the wave function changes when the space is enlarged, the couplings are recalculated after diagonalization in $\mathcal V'$.

The iteration starts from configurations that contain the principal components of the desired physical states.
As shown in Fig.~\ref{fig:chci}(a), diagonalization, evaluation of $S_G$, and expansion of the variational space are repeated until no new configuration has $S_G\geq\epsilon$.
For single-state selection, $\mathcal K$ contains one target level and the selected space is tailored to that state.
For multistate selection, $\mathcal K$ contains several levels of the same symmetry and the maximum in Eq.~\eqref{eq:configuration-selection} builds a shared space containing configurations important to any of them.
Both choices use the same selection rule and leave the remaining CSFs for the energy correction.

After the variational space has converged, we evaluate an Epstein--Nesbet second-order correction~\cite{Sharma2017SHCI,Holmes2017ExcitedHCI}.
The zeroth-order Hamiltonian retains the full matrix within $\mathcal V$ and the diagonal elements $H_{aa}=\langle\Phi_a|H_{\mathrm{eff}}|\Phi_a\rangle$ in the external space.
The second-order energy correction for target state $k$ is
\begin{equation}
 E_k^{(2)}=\sum_{a\in\mathcal C\setminus\mathcal V}
     \frac{|\sigma_{ak}|^2}{E_k^{\mathrm{var}}-H_{aa}}.
\label{eq:en-correction}
\end{equation}
The coupling magnitude and the energy separation together determine the contribution of each external CSF.
The corrected energy is then
\begin{equation}
 E_k^{\mathrm{tot}}=E_k^{\mathrm{var}}+E_k^{(2)}.
 \label{eq:corrected-energy}
\end{equation}
Core excitations enter through $H_{\mathrm{eff}}$, while Eq.~\eqref{eq:en-correction} accounts for valence correlation outside the selected variational space.
This final step is indicated by the energy-correction box in Fig.~\ref{fig:chci}(a).

In each $J^\pi$ block, the initial configurations contain the principal components of the target states.
Their continuity during space expansion is monitored through wave-function overlaps between successive iterations, evaluated by matching the shared CSFs.
Comparing overlaps with the computed roots reveals possible changes in state ordering.
Only the Hamiltonian within $\mathcal V$ is diagonalized.
Couplings to external CSFs are evaluated on demand in blocks over connected configurations and contracted with the variational coefficients to obtain $\sigma_{ak}$ for selection and the second-order correction.
The full external-to-variational coupling matrix therefore need not be stored in memory.
\section{Results}
\label{sec:results}

We perform CI--MBPT and CHCI--MBPT calculations for neutral Al, Si, and P.
The CI--MBPT results serve as benchmarks for assessing the reliability of CHCI--MBPT and its ability to compress the CI space.
We first examine the dependence on the selection threshold in Al, then extend the comparison to Si and P, which have more valence electrons.
We then relate the differences in compression to the configuration mixing of the target states and the composition of the parent CI spaces.

For all three atoms, the one-electron basis is constructed using 40 B-splines of order $k=8$ in a cavity of radius $20a_0$, with a closed-shell $[\mathrm{Ne}]$ core.

\subsection{Al}
\label{sec:results-al}

Al provides a three-valence-electron system in which to examine the threshold dependence of CHCI--MBPT.
The reference configurations are $3s^23p$, $3s^24p$, $3s^24s$, $3s^23d$, and $3s3p^2$.
Allowing excitations of all three valence electrons into the $18spdfg$ orbital set generates the full CI space.
In the MBPT calculations, the self-energy operator is evaluated using the $25spdfgh$ orbital set.
We calculate two levels in each of the odd-parity $J=1/2,3/2$ blocks and the even-parity $J=1/2,3/2,5/2$ blocks, giving ten target levels.
The CI dimensions range from 274750 to 624580 CSFs.

\begingroup
\expandafter\let\expandafter\table\csname table*\endcsname
\expandafter\let\expandafter\endtable\csname endtable*\endcsname
\begin{table}[!htbp]
\caption{CHCI--MBPT results for Al at $\epsilon=2\times10^{-4}\,E_{\mathrm h}$. $\delta E$ is the excitation-energy difference between CHCI--MBPT and CI--MBPT. $E^{(2)}$ gives the second-order correction to the ground-state total energy or to the excitation energy for excited states. $C=N_{\mathrm{CI}}/N_{\mathrm{var}}$ is the CSF-space compression factor. Energies are in $\mathrm{cm}^{-1}$. Subscripts m and s denote multistate and single-state selection.}
\label{tab:al}
\small
\setlength{\tabcolsep}{3pt}
\begin{ruledtabular}
\begin{tabular}{lrrrrrrrr}
 & \multicolumn{5}{c}{Energy ($\mathrm{cm}^{-1}$)} & \multicolumn{3}{c}{CI space} \\ \cline{2-6}\cline{7-9}
State & $E_{\rm CI}$ & $\delta E_{\rm m}$ & $\delta E_{\rm s}$ & $E^{(2)}_{\rm m}$ & $E^{(2)}_{\rm s}$ & $N_{\rm CI}$ & $C_{\rm m}$ & $C_{\rm s}$ \\ \hline
$3s^2 3p\;{}^{2}P^{\mathrm{o}}_{1/2}$ & 0.00 & +0.00 & +0.00 & $-$2.47 & $-$3.19 & 274750 & 15.52 & 19.87 \\
$3s^2 3p\;{}^{2}P^{\mathrm{o}}_{3/2}$ & 110.84 & $-$0.29 & $-$0.35 & $-$1.67 & $-$1.83 & 494410 & 18.91 & 24.85 \\
$3s^2 4s\;{}^{2}S_{1/2}$ & 25368.36 & +0.14 & +0.09 & +0.60 & +0.37 & 275676 & 16.85 & 29.36 \\
$3s3p^2\;{}^{4}P_{1/2}$ & 28913.27 & $-$0.28 & $-$0.30 & $-$1.97 & $-$2.12 & 275676 & 16.85 & 20.59 \\
$3s3p^2\;{}^{4}P_{3/2}$ & 28959.59 & $-$0.44 & $-$0.45 & $-$2.67 & $-$2.87 & 495764 & 17.00 & 31.07 \\
$3s3p^2\;{}^{4}P_{5/2}$ & 29034.85 & $-$0.49 & $-$0.52 & $-$3.19 & $-$3.58 & 624580 & 19.19 & 36.87 \\
$3s^2 3d\;{}^{2}D_{3/2}$ & 32435.08 & $-$0.40 & $-$0.53 & $-$2.70 & $-$3.37 & 495764 & 17.00 & 20.27 \\
$3s^2 3d\;{}^{2}D_{5/2}$ & 32436.49 & $-$0.56 & $-$0.74 & $-$3.34 & $-$4.12 & 624580 & 19.19 & 21.66 \\
$3s^2 4p\;{}^{2}P^{\mathrm{o}}_{1/2}$ & 33157.16 & +0.22 & +0.10 & +0.95 & +0.59 & 274750 & 15.52 & 21.67 \\
$3s^2 4p\;{}^{2}P^{\mathrm{o}}_{3/2}$ & 33174.23 & +0.08 & $-$0.12 & +0.11 & $-$0.80 & 494410 & 18.91 & 27.08 \\
\end{tabular}
\end{ruledtabular}
\end{table}
\endgroup

\begin{figure}[t]
\centering
\includegraphics[width=\columnwidth]{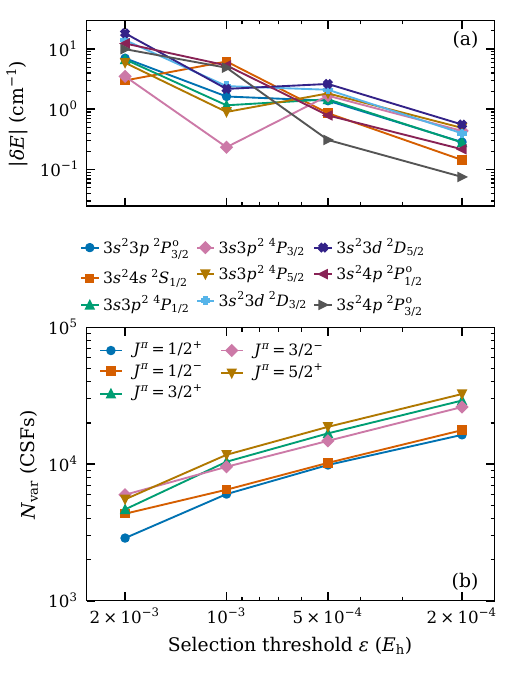}
\caption{Threshold dependence of multistate CHCI--MBPT for Al.
(a) Absolute excitation-energy differences from CI--MBPT.
(b) Number of variational CSFs in each $J^\pi$ block.}
\label{fig:threshold}
\end{figure}

Figure~\ref{fig:threshold}(a) compares multistate CHCI--MBPT with the CI--MBPT benchmark at $\epsilon=2\times10^{-3}$, $10^{-3}$, $5\times10^{-4}$, and $2\times10^{-4}E_{\mathrm h}$, where $E_{\mathrm h}$ is the Hartree energy.
The excitation energies $E$ are measured relative to the ground state obtained with the same method, and their differences are denoted by $\delta E_{k,\mathrm m}=E_{k,\mathrm m}-E_{k,\mathrm{CI}}$, with m indicating multistate selection.
As the threshold is lowered, the maximum absolute difference among the nine excited levels decreases from 18.22 to 6.21, 2.63, and $0.56\ \mathrm{cm}^{-1}$, respectively.
The root-mean-square difference over these levels decreases from 10.18 to $0.36\ \mathrm{cm}^{-1}$.
The approximately linear trends on the double-logarithmic axes motivate power-law fits to the threshold dependence.
Fits over the four thresholds show that the maximum absolute excitation-energy difference scales as $\epsilon^{1.49}$, while the root-mean-square difference scales as $\epsilon^{1.43}$.
These fits describe the overall improvement in the spectrum despite local nonmonotonic changes in individual excitation-energy differences.
The CHCI--MBPT total energies of Al I and Al II are also well described by a power-law dependence on the threshold over the sampled interval, as shown in Appendix~\ref{app:al-total-energies}.

\begin{table}[t]
\caption{Multistate CHCI--MBPT results for Al at $\epsilon=2\times10^{-3}\,E_{\mathrm h}$. $\delta E$ is the excitation-energy difference between CHCI--MBPT and CI--MBPT. $E^{(2)}$ gives the second-order correction to the ground-state total energy or to the excitation energy for excited states. $C=N_{\mathrm{CI}}/N_{\mathrm{var}}$ is the CSF-space compression factor. Energies are in $\mathrm{cm}^{-1}$.}
\label{tab:al-loose}
\small
\setlength{\tabcolsep}{5pt}
\begin{ruledtabular}
\begin{tabular}{lrrr}
 & \multicolumn{2}{c}{Energy ($\mathrm{cm}^{-1}$)} & CI space \\ \cline{2-3}\cline{4-4}
State & $\delta E$ & $E^{(2)}$ & $C$ \\ \hline
$3s^2 3p\;{}^{2}P^{\mathrm{o}}_{1/2}$ & +0.00 & $-$99.32 & 63.42 \\
$3s^2 3p\;{}^{2}P^{\mathrm{o}}_{3/2}$ & $-$7.02 & $-$31.95 & 82.90 \\
$3s^2 4s\;{}^{2}S_{1/2}$ & +3.00 & $-$5.74 & 95.92 \\
$3s3p^2\;{}^{4}P_{1/2}$ & $-$6.72 & $-$51.76 & 95.92 \\
$3s3p^2\;{}^{4}P_{3/2}$ & $-$3.51 & $-$33.93 & 106.41 \\
$3s3p^2\;{}^{4}P_{5/2}$ & $-$5.91 & $-$43.80 & 113.15 \\
$3s^2 3d\;{}^{2}D_{3/2}$ & $-$14.07 & $-$90.80 & 106.41 \\
$3s^2 3d\;{}^{2}D_{5/2}$ & $-$18.22 & $-$103.70 & 113.15 \\
$3s^2 4p\;{}^{2}P^{\mathrm{o}}_{1/2}$ & +12.36 & +40.04 & 63.42 \\
$3s^2 4p\;{}^{2}P^{\mathrm{o}}_{3/2}$ & +9.94 & +21.19 & 82.90 \\
\end{tabular}
\end{ruledtabular}
\end{table}

The accompanying growth of the variational spaces is shown in Fig.~\ref{fig:threshold}(b).
Across the five symmetry blocks, fits to $N_{\mathrm{var}}\propto\epsilon^p$ give exponents $p$ between $-0.78$ and $-0.62$.
The sum of the five block dimensions, denoted by $N_{\mathrm{var,tot}}$, follows $N_{\mathrm{var,tot}}\propto\epsilon^{-0.71}$ over the sampled interval.
Reducing the threshold by a factor of ten increases the total variational dimension from 23349 to 121914 CSFs, a factor of 5.22.
We use the stringent threshold $\epsilon=2\times10^{-4}E_{\mathrm h}$ for the comparison of multistate and single-state CHCI--MBPT in Table~\ref{tab:al}.

The configuration-selection cycle converges in a small number of iterations.
Here, an iteration comprises diagonalization in the current variational space
followed by configuration selection, and the count includes the final cycle
in which no new configurations are added.
For multistate selection across the four Al thresholds, from $2\times10^{-3}$ to
$2\times10^{-4}E_{\mathrm h}$, the five $J^\pi$ blocks require three to five
iterations.
At $\epsilon=2\times10^{-4}E_{\mathrm h}$, all five blocks converge in four
iterations.
Single-state selection at this threshold requires four to five iterations.
The iteration count does not increase monotonically as the threshold is
lowered: the $J^\pi=3/2^+$ block requires five iterations at
$10^{-3}E_{\mathrm h}$ and four at both $5\times10^{-4}$ and
$2\times10^{-4}E_{\mathrm h}$.
Thus, the increase in the variational dimensions at smaller thresholds is
accompanied by only a modest variation in the number of updates of the wave
functions and selected configurations.

Table~\ref{tab:al} compares the excitation energies and space compression at this threshold, with subscripts m and s denoting multistate and single-state selection.
The compression factor is $C_X=N_{\mathrm{CI}}/N_{\mathrm{var},X}$ for $X\in\{\mathrm m,\mathrm s\}$.
Multistate selection reduces the CI dimensions by factors of 15.52--19.19, while single-state selection gives factors of 19.87--36.87.
This threshold already gives excitation-energy differences below $10.00\ \mathrm{cm}^{-1}$ without the EN correction, with maxima of 2.78 and $3.38\ \mathrm{cm}^{-1}$ for multistate and single-state selection, respectively.
Including the correction brings most differences below $0.50\ \mathrm{cm}^{-1}$, with corresponding maxima of 0.56 and $0.74\ \mathrm{cm}^{-1}$.
For the $3s3p^2\,{}^4P_{5/2}$ level, the compression factor increases from 19.19 to 36.87 when changing from multistate to single-state selection, while the energy difference changes from $-0.49$ to $-0.52\ \mathrm{cm}^{-1}$.
Thus, a space selected for one target level can be substantially smaller than the space shared by two levels while retaining similar excitation-energy accuracy after the external-space correction.

\begingroup
\expandafter\let\expandafter\table\csname table*\endcsname
\expandafter\let\expandafter\endtable\csname endtable*\endcsname
\begin{table}[!htbp]
\caption{CHCI--MBPT results for Si at $\epsilon=2\times10^{-4}\,E_{\mathrm h}$. $\delta E$ is the excitation-energy difference between CHCI--MBPT and CI--MBPT. $E^{(2)}$ gives the second-order correction to the ground-state total energy or to the excitation energy for excited states. $C=N_{\mathrm{CI}}/N_{\mathrm{var}}$ is the CSF-space compression factor. Energies are in $\mathrm{cm}^{-1}$. Subscripts m and s denote multistate and single-state selection.}
\label{tab:si}
\small
\setlength{\tabcolsep}{3pt}
\begin{ruledtabular}
\begin{tabular}{lrrrrrrrr}
 & \multicolumn{5}{c}{Energy ($\mathrm{cm}^{-1}$)} & \multicolumn{3}{c}{CI space} \\ \cline{2-6}\cline{7-9}
State & $E_{\rm CI}$ & $\delta E_{\rm m}$ & $\delta E_{\rm s}$ & $E^{(2)}_{\rm m}$ & $E^{(2)}_{\rm s}$ & $N_{\rm CI}$ & $C_{\rm m}$ & $C_{\rm s}$ \\ \hline
$3s^2 3p^2\;{}^{3}P_{0}$ & 0.00 & +0.00 & +0.00 & $-$1.59 & $-$3.06 & 58060 & 1.91 & 2.19 \\
$3s^2 3p^2\;{}^{3}P_{1}$ & 75.97 & $-$0.29 & $-$0.37 & $-$1.64 & $-$2.16 & 158691 & 1.97 & 2.70 \\
$3s^2 3p^2\;{}^{3}P_{2}$ & 220.60 & $-$0.47 & $-$0.63 & $-$2.68 & $-$3.56 & 223557 & 2.32 & 2.61 \\
$3s^2 3p^2\;{}^{1}D_{2}$ & 6481.53 & $-$1.10 & $-$1.00 & $-$5.66 & $-$5.15 & 223557 & 2.32 & 2.41 \\
$3s^2 3p^2\;{}^{1}S_{0}$ & 16306.00 & $-$0.25 & $-$0.05 & $-$1.42 & $-$0.49 & 58060 & 1.91 & 2.00 \\
$3s3p^3\;{}^{5}S^{\mathrm{o}}_{2}$ & 33421.97 & +0.04 & +0.32 & +0.18 & +1.04 & 134736 & 1.90 & 2.73 \\
$3s^2 3p4s\;{}^{3}P^{\mathrm{o}}_{0}$ & 40400.30 & +0.08 & +0.03 & +0.55 & +0.47 & 35122 & 1.50 & 2.45 \\
$3s^2 3p4s\;{}^{3}P^{\mathrm{o}}_{1}$ & 40478.03 & $-$0.38 & $-$0.31 & $-$1.50 & $-$1.24 & 96588 & 2.20 & 2.37 \\
$3s^2 3p4s\;{}^{3}P^{\mathrm{o}}_{2}$ & 40668.45 & $-$0.24 & $-$0.66 & $-$1.19 & $-$2.89 & 134736 & 1.90 & 3.46 \\
$3s^2 3p4s\;{}^{1}P^{\mathrm{o}}_{1}$ & 41860.18 & $-$0.40 & $-$0.28 & $-$1.60 & $-$1.11 & 96588 & 2.20 & 2.36 \\
$3s^2 3p4p\;{}^{1}P_{1}$ & 48333.35 & $-$0.38 & $-$0.20 & $-$2.21 & $-$1.54 & 158691 & 1.97 & 2.34 \\
$3s^2 3p3d\;{}^{3}P^{\mathrm{o}}_{0}$ & 52057.26 & $-$0.02 & +0.11 & +0.18 & +0.65 & 35122 & 1.50 & 1.93 \\
\end{tabular}
\end{ruledtabular}
\end{table}
\endgroup

\begin{table}[!htbp]
\caption{Multistate CHCI--MBPT results for Si at $\epsilon=1.5\times10^{-3}\,E_{\mathrm h}$. $\delta E$ is the excitation-energy difference between CHCI--MBPT and CI--MBPT. $E^{(2)}$ gives the second-order correction to the ground-state total energy or to the excitation energy for excited states. $C=N_{\mathrm{CI}}/N_{\mathrm{var}}$ is the CSF-space compression factor. Energies are in $\mathrm{cm}^{-1}$.}
\label{tab:si-loose}
\small
\setlength{\tabcolsep}{3pt}
\begin{ruledtabular}
\begin{tabular}{lrrr}
 & \multicolumn{2}{c}{Energy ($\mathrm{cm}^{-1}$)} & CI space \\ \cline{2-3}\cline{4-4}
State & $\delta E$ & $E^{(2)}$ & $C$ \\ \hline
$3s^2 3p^2\;{}^{3}P_{0}$ & +0.00 & $-$82.89 & 5.38 \\
$3s^2 3p^2\;{}^{3}P_{1}$ & $-$19.01 & $-$78.81 & 5.72 \\
$3s^2 3p^2\;{}^{3}P_{2}$ & $-$23.90 & $-$104.22 & 8.38 \\
$3s^2 3p^2\;{}^{1}D_{2}$ & $-$45.10 & $-$188.88 & 8.38 \\
$3s^2 3p^2\;{}^{1}S_{0}$ & $-$14.70 & $-$56.00 & 5.38 \\
$3s3p^3\;{}^{5}S^{\mathrm{o}}_{2}$ & $-$18.86 & $-$56.41 & 6.58 \\
$3s^2 3p4s\;{}^{3}P^{\mathrm{o}}_{0}$ & +7.53 & +40.24 & 3.23 \\
$3s^2 3p4s\;{}^{3}P^{\mathrm{o}}_{1}$ & $-$1.49 & $-$7.76 & 6.23 \\
$3s^2 3p4s\;{}^{3}P^{\mathrm{o}}_{2}$ & $-$1.87 & $-$14.45 & 6.58 \\
$3s^2 3p4s\;{}^{1}P^{\mathrm{o}}_{1}$ & $-$0.45 & $-$7.93 & 6.23 \\
$3s^2 3p4p\;{}^{1}P_{1}$ & $-$2.51 & $-$37.62 & 5.72 \\
$3s^2 3p3d\;{}^{3}P^{\mathrm{o}}_{0}$ & +1.97 & +16.71 & 3.23 \\
\end{tabular}
\end{ruledtabular}
\end{table}

The role of the EN correction becomes more pronounced at looser thresholds, where more of the correlation energy lies outside the variational space.
Table~\ref{tab:al-loose} lists the multistate results at $\epsilon=2\times10^{-3}E_{\mathrm h}$ for the same ten levels.
In the tables, the $E^{(2)}$ columns give the ground-state total-energy correction for the ground state and the difference $E_k^{(2)}-E_0^{(2)}$ for each excited state, where 0 labels the ground state.
At $\epsilon=2\times10^{-3}E_{\mathrm h}$, multistate selection compresses the CI spaces by factors of 63.42--113.15, substantially exceeding those at the stringent threshold.
The maximum excitation-energy difference at this looser threshold is $85.48\ \mathrm{cm}^{-1}$ before the EN correction and $18.22\ \mathrm{cm}^{-1}$ after it.

\subsection{Si and P}
\label{sec:results-si-p}

We next apply CHCI--MBPT to Si and P, which have four and five valence electrons, respectively.
For both atoms, we compare single-state and multistate selection at $\epsilon=2\times10^{-4}E_{\mathrm h}$ and examine stronger compression with multistate selection at $\epsilon=1.5\times10^{-3}E_{\mathrm h}$.
In the MBPT calculations for both atoms, the self-energy operator is evaluated using the $30spdfgh$ orbital set.

\begingroup
\expandafter\let\expandafter\table\csname table*\endcsname
\expandafter\let\expandafter\endtable\csname endtable*\endcsname
\begin{table}[!htbp]
\caption{CHCI--MBPT results for P at $\epsilon=2\times10^{-4}\,E_{\mathrm h}$. $\delta E$ is the excitation-energy difference between CHCI--MBPT and CI--MBPT. $E^{(2)}$ gives the second-order correction to the ground-state total energy or to the excitation energy for excited states. $C=N_{\mathrm{CI}}/N_{\mathrm{var}}$ is the CSF-space compression factor. Energies are in $\mathrm{cm}^{-1}$. Subscripts m and s denote multistate and single-state selection.}
\label{tab:p}
\small
\setlength{\tabcolsep}{3pt}
\begin{ruledtabular}
\begin{tabular}{lrrrrrrrr}
 & \multicolumn{5}{c}{Energy ($\mathrm{cm}^{-1}$)} & \multicolumn{3}{c}{CI space} \\ \cline{2-6}\cline{7-9}
State & $E_{\rm CI}$ & $\delta E_{\rm m}$ & $\delta E_{\rm s}$ & $E^{(2)}_{\rm m}$ & $E^{(2)}_{\rm s}$ & $N_{\rm CI}$ & $C_{\rm m}$ & $C_{\rm s}$ \\ \hline
$3s^2 3p^3\;{}^{4}S^{\mathrm{o}}_{3/2}$ & 0.00 & +0.00 & +0.00 & $-$5.76 & $-$10.11 & 406651 & 1.71 & 2.01 \\
$3s^2 3p^3\;{}^{2}D^{\mathrm{o}}_{3/2}$ & 11805.96 & $-$0.83 & $-$0.77 & $-$3.23 & $-$3.18 & 406651 & 1.71 & 1.88 \\
$3s^2 3p^3\;{}^{2}D^{\mathrm{o}}_{5/2}$ & 11820.42 & $-$1.22 & $-$1.92 & $-$4.46 & $-$7.03 & 496207 & 1.89 & 2.51 \\
$3s^2 3p^3\;{}^{2}P^{\mathrm{o}}_{1/2}$ & 19942.12 & $-$0.29 & $-$0.61 & $-$0.76 & $-$1.43 & 229707 & 1.53 & 2.02 \\
$3s^2 3p^3\;{}^{2}P^{\mathrm{o}}_{3/2}$ & 19963.43 & $-$0.93 & $-$1.40 & $-$3.75 & $-$5.30 & 406651 & 1.71 & 2.08 \\
$3s^2 3p^2({}^3P)4s\;{}^{4}P_{1/2}$ & 58930.69 & +0.87 & +1.31 & +4.38 & +6.21 & 123215 & 1.24 & 1.66 \\
$3s^2 3p^2({}^3P)4s\;{}^{4}P_{3/2}$ & 59080.76 & +0.65 & +0.61 & +3.35 & +3.16 & 216741 & 1.44 & 1.99 \\
$3s^2 3p^2({}^3P)4s\;{}^{4}P_{5/2}$ & 59327.39 & +0.63 & +0.70 & +3.19 & +3.75 & 262567 & 1.43 & 1.79 \\
$3s^2 3p^2({}^3P)4s\;{}^{2}P_{1/2}$ & 61513.42 & +0.86 & +1.62 & +4.38 & +7.68 & 123215 & 1.24 & 1.34 \\
$3s^2 3p^2({}^3P)4s\;{}^{2}P_{3/2}$ & 61810.57 & +0.63 & +1.30 & +3.32 & +6.14 & 216741 & 1.44 & 1.52 \\
$3s3p^4\;{}^{4}P_{5/2}$ & 63119.99 & $-$0.47 & $-$0.03 & $-$1.21 & +0.65 & 262567 & 1.43 & 1.59 \\
$3s3p^4\;{}^{4}P_{3/2}$ & 63306.83 & $-$0.66 & $-$0.25 & $-$2.00 & $-$0.06 & 216741 & 1.44 & 1.51 \\
$3s3p^4\;{}^{4}P_{1/2}$ & 63413.93 & +0.52 & +0.48 & +2.89 & +3.11 & 123215 & 1.24 & 1.70 \\
$3s^2 3p^2({}^3P)4p\;{}^{2}S^{\mathrm{o}}_{1/2}$ & 67051.94 & +0.49 & +0.85 & +2.37 & +3.40 & 229707 & 1.53 & 1.83 \\
$3s^2 3p^2({}^3P)4p\;{}^{4}D^{\mathrm{o}}_{1/2}$ & 68421.45 & +0.60 & +0.77 & +2.58 & +2.62 & 229707 & 1.53 & 1.94 \\
$3s^2 3p^2({}^3P)4p\;{}^{4}D^{\mathrm{o}}_{5/2}$ & 68632.60 & +0.08 & $-$0.03 & +0.18 & $-$1.78 & 496207 & 1.89 & 2.34 \\
$3s^2 3p^2({}^1D)4s\;{}^{2}D_{5/2}$ & 69247.25 & +0.07 & $-$0.03 & +0.96 & +0.06 & 262567 & 1.43 & 1.66 \\
$3s^2 3p^2({}^3P)4p\;{}^{4}P^{\mathrm{o}}_{5/2}$ & 69722.34 & $-$0.03 & +0.05 & $-$0.01 & $-$0.77 & 496207 & 1.89 & 2.24 \\
\end{tabular}
\end{ruledtabular}
\end{table}
\endgroup

\begin{table}[!htbp]
\caption{Multistate CHCI--MBPT results for P at $\epsilon=1.5\times10^{-3}\,E_{\mathrm h}$. $\delta E$ is the excitation-energy difference between CHCI--MBPT and CI--MBPT. $E^{(2)}$ gives the second-order correction to the ground-state total energy or to the excitation energy for excited states. $C=N_{\mathrm{CI}}/N_{\mathrm{var}}$ is the CSF-space compression factor. Energies are in $\mathrm{cm}^{-1}$.}
\label{tab:p-loose}
\small
\setlength{\tabcolsep}{3pt}
\begin{ruledtabular}
\begin{tabular}{lrrr}
 & \multicolumn{2}{c}{Energy ($\mathrm{cm}^{-1}$)} & CI space \\ \cline{2-3}\cline{4-4}
State & $\delta E$ & $E^{(2)}$ & $C$ \\ \hline
$3s^2 3p^3\;{}^{4}S^{\mathrm{o}}_{3/2}$ & +0.00 & $-$382.24 & 6.78 \\
$3s^2 3p^3\;{}^{2}D^{\mathrm{o}}_{3/2}$ & $-$38.34 & $-$119.51 & 6.78 \\
$3s^2 3p^3\;{}^{2}D^{\mathrm{o}}_{5/2}$ & $-$17.60 & $-$36.37 & 5.73 \\
$3s^2 3p^3\;{}^{2}P^{\mathrm{o}}_{1/2}$ & +12.41 & +77.68 & 3.87 \\
$3s^2 3p^3\;{}^{2}P^{\mathrm{o}}_{3/2}$ & $-$39.85 & $-$113.80 & 6.78 \\
$3s^2 3p^2({}^3P)4s\;{}^{4}P_{1/2}$ & +67.21 & +290.72 & 2.65 \\
$3s^2 3p^2({}^3P)4s\;{}^{4}P_{3/2}$ & +60.59 & +258.80 & 4.02 \\
$3s^2 3p^2({}^3P)4s\;{}^{4}P_{5/2}$ & +59.22 & +255.93 & 3.57 \\
$3s^2 3p^2({}^3P)4s\;{}^{2}P_{1/2}$ & +68.65 & +291.28 & 2.65 \\
$3s^2 3p^2({}^3P)4s\;{}^{2}P_{3/2}$ & +63.12 & +265.31 & 4.02 \\
$3s3p^4\;{}^{4}P_{5/2}$ & +0.68 & +57.95 & 3.57 \\
$3s3p^4\;{}^{4}P_{3/2}$ & $-$44.35 & $-$92.70 & 4.02 \\
$3s3p^4\;{}^{4}P_{1/2}$ & +35.12 & +186.03 & 2.65 \\
$3s^2 3p^2({}^3P)4p\;{}^{2}S^{\mathrm{o}}_{1/2}$ & +61.66 & +220.48 & 3.87 \\
$3s^2 3p^2({}^3P)4p\;{}^{4}D^{\mathrm{o}}_{1/2}$ & +64.97 & +230.00 & 3.87 \\
$3s^2 3p^2({}^3P)4p\;{}^{4}D^{\mathrm{o}}_{5/2}$ & +56.17 & +173.32 & 5.73 \\
$3s^2 3p^2({}^1D)4s\;{}^{2}D_{5/2}$ & +43.00 & +183.18 & 3.57 \\
$3s^2 3p^2({}^3P)4p\;{}^{4}P^{\mathrm{o}}_{5/2}$ & +56.28 & +177.36 & 5.73 \\
\end{tabular}
\end{ruledtabular}
\end{table}

For Si, the reference configurations are $3s^23p^2$ and $3s^23p4p$ for even parity, and $3s^23p4s$ and $3s3p^3$ for odd parity.
Single and double excitations from these references into the $18spdfg$ orbital set define the CI spaces.
We calculate two levels in each $J=0,1,2$ block of both parities, giving twelve target levels and CI dimensions ranging from 35122 to 223557 CSFs.

At the stringent threshold, Table~\ref{tab:si} gives compression factors of 1.50--2.32 for multistate selection and 1.93--3.46 for single-state selection.
The maximum absolute excitation-energy differences from CI--MBPT are 1.10 and $1.00\ \mathrm{cm}^{-1}$, respectively, after the EN correction.

Raising the threshold to $1.5\times10^{-3}E_{\mathrm h}$ increases the multistate compression factors to 3.23--8.38, as shown in Table~\ref{tab:si-loose}.
The largest absolute excitation-energy deviation occurs for $3s^23p^2\,{}^1D_2$ both before and after the correction.
For this level, the differential correction $E_k^{(2)}-E_0^{(2)}=-188.88\ \mathrm{cm}^{-1}$ changes the excitation-energy deviation from $+143.78$ to $-45.10\ \mathrm{cm}^{-1}$.

For P, the odd-parity reference configurations are $3s^23p^3$ and $3s^23p^24p$, and the even-parity references are $3s^23p^24s$ and $3s3p^4$.
The CI spaces include single excitations into $18spdfg$ and double excitations into $16spdfg$ from the two references of each parity.
We calculate three levels in each $J=1/2,3/2,5/2$ block of both parities, giving eighteen target levels.
The corresponding CI dimensions range from 123215 to 496207 CSFs.

At $\epsilon=2\times10^{-4}E_{\mathrm h}$, the P results in Table~\ref{tab:p} give compression factors of 1.24--1.89 for multistate selection and 1.34--2.51 for single-state selection.
The maximum absolute excitation-energy differences are 1.22 and $1.92\ \mathrm{cm}^{-1}$, respectively.

At the looser threshold, the multistate compression factors increase to 2.65--6.78 (Table~\ref{tab:p-loose}).
The maximum absolute excitation-energy difference decreases from 223.50 to $68.65\ \mathrm{cm}^{-1}$ when the EN correction is included.

The ground-state corrections in Tables~\ref{tab:si-loose} and \ref{tab:p-loose} show a larger external-space contribution for P than for Si.
For example, the ground-state correction is $-382.24\ \mathrm{cm}^{-1}$ for P, compared with $-82.89\ \mathrm{cm}^{-1}$ for Si at the same threshold.
Both parent spaces are constructed from single and double excitations of two reference configurations for each parity, so this comparison probes the residual correlation within the same type of excitation truncation.
The threshold controls individual configuration-coupling amplitudes, whereas Eq.~\eqref{eq:en-correction} sums the energy contributions of all external CSFs, weighted by their energy denominators.
Many weak couplings can therefore yield a substantial accumulated correction.
The larger correction for P indicates that more correlation energy remains outside its selected variational space, consistent with the additional electron-pair correlation channels available in the five-valence-electron system.

The excitation-energy corrections among the P levels also reveal a dependence on orbital occupation.
For the excited $3s^23p^3$ levels, these corrections range from $-119.51$ to $+77.68\ \mathrm{cm}^{-1}$, whereas the $3s^23p^2({}^3P)4s$ levels receive positive corrections of $255.93$--$291.28\ \mathrm{cm}^{-1}$.
The latter arise because the EN correction lowers the ground state more than these $4s$ excited states, increasing their excitation energies.
For $3s^23p^2({}^3P)4s\,{}^2P_{1/2}$, the differential EN correction is $+291.28\ \mathrm{cm}^{-1}$, offsetting a variational deviation of $-222.64\ \mathrm{cm}^{-1}$ and leaving $+68.65\ \mathrm{cm}^{-1}$.
For the Si ${}^1D_2$ level discussed above, the excited state instead receives the larger downward correction, reducing its excitation energy.
Thus, the accuracy of the excitation energies depends on recovering the different residual correlation contributions associated with the ground and excited states.

The configuration-selection cycle converges in a few iterations for both atoms.
At the stringent threshold, Si requires four to five iterations for both selection modes, while P requires four to five for multistate selection and four to seven for single-state selection.
At $\epsilon=1.5\times10^{-3}E_{\mathrm h}$, multistate selection requires four to seven iterations for each atom.
The number of updates remains comparable at the two thresholds despite the reduction in variational dimension.
The few selection cycles required here are also comparable in number to the five to six refinement rounds reported in atomic neural-network selection studies~\cite{Bilous2023GRASP,Bilous2024AtomicNN}, although the selection thresholds and stopping conditions differ.
\subsection{Correlation structure and CI-space compression}
\label{sec:results-discussion}

The differences in compression across Al, Si, and P reflect both the configuration mixing of the target states and the composition of the parent CI spaces.
In the CI--MBPT benchmark wave functions, the leading nonrelativistic configurations of the ground states, $3s^23p$, $3s^23p^2$, and $3s^23p^3$, carry approximately 77\%, 66\%, and 55\% of the norm, respectively.
The increasing contribution from other configurations indicates stronger configuration mixing along this sequence in the orbital bases used here.
Such mixing requires an appreciable set of configurations to remain in the variational space and is consistent with the smaller compression factors for Si and P.

The computational demands also differ substantially despite comparable CI dimensions.
The largest CI blocks contain approximately $6.25\times10^5$ CSFs for Al and $4.96\times10^5$ for P, whereas the corresponding benchmark calculations reach peak memory usages of approximately 573 and 1686 GiB, respectively.
Together with the stronger configuration mixing in P, this contrast highlights that the complexity of the many-electron calculation is not captured by the CI dimension alone.

The relationship between single-state and multistate compression depends on how much of this correlation space is shared by the target levels.
For example, the Si $3s^23p^2\,{}^3P_2$ and ${}^1D_2$ levels have the same dominant nonrelativistic configuration, as do the three low-lying P levels in the odd-parity $J=3/2$ block.
Their common orbital occupations provide similar radial correlation channels, so configurations important to one level can also contribute to the others.
Multistate selection can therefore accommodate these levels with relatively little additional space compared with single-state selection.
In contrast, the Al $3s^24s\,{}^2S_{1/2}$ and $3s3p^2\,{}^4P_{1/2}$ levels involve different orbital occupations and require different sets of correlation configurations.
A space tailored to either level can omit configurations needed primarily by the other, giving a more pronounced advantage for single-state selection.
The selection criterion in Eq.~\eqref{eq:configuration-selection} expresses this distinction through the coherent couplings to the target wave functions.

The parent spaces provide a second reason for the larger compression in Al.
Its three-electron full CI space extends to $18spdfg$ and contains many configurations involving high-lying virtual orbitals that couple weakly to the low-lying target states.
The threshold excludes these weakly coupled configurations from the variational space, producing the large reductions seen in Tables~\ref{tab:al} and \ref{tab:al-loose}.
The Si and P spaces already focus on single and double excitations from two reference configurations for each parity.
Together with the stronger configuration mixing, this restriction leaves a larger fraction of the parent space relevant to the target states, and hence a smaller compression factor at the same threshold.

The effectiveness of a perturbative external-space treatment depends on how this partition is made.
CIPT incorporates external configurations through an effective interaction, whereas emu CI retains their coupling to the important subspace in the diagonalization~\cite{Dzuba2017CIPT,Geddes2018EmuCI}.
The nonperturbative treatment in emu CI accounts for the feedback of these configurations on the eigenstates.
CHCI adapts the partition to the coherent couplings of configurations to the target wave functions.
Strongly coupled configurations enter the variational diagonalization, and the EN correction recovers the accumulated contribution of the remaining weak couplings.
The improvement in the calculated energies demonstrates that this adaptive partition leaves an external space that can be treated effectively at second order for the present systems.

The EN correction brings the total energies of all calculated states in Tables~\ref{tab:al}--\ref{tab:p-loose} closer to their CI--MBPT benchmarks.
Across these results, the mean absolute total-energy deviation decreases from 50.84 to $14.09\ \mathrm{cm}^{-1}$, a reduction by a factor of 3.61.
For the excitation energies, the corresponding mean absolute deviation decreases from 28.84 to $9.40\ \mathrm{cm}^{-1}$, a factor of 3.07.
Only three excitation energies at the stringent threshold show a small increase in absolute deviation, in each case below $0.07\ \mathrm{cm}^{-1}$, with final absolute deviations no greater than $0.25\ \mathrm{cm}^{-1}$.
These small exceptions arise from the cancellation between ground- and excited-state errors, even though both total energies improve.
For every atom, threshold, and selection mode, the maximum and root-mean-square excitation-energy deviations are reduced.
Further refinement can proceed through a sequence of decreasing selection thresholds, with the resulting energies also available for extrapolation~\cite{Holmes2016HCI,Sharma2017SHCI}.
Appendix~\ref{app:al-total-energies} examines this threshold dependence for the Al I and Al II total energies and illustrates the extraction of zero-threshold energies.
By recovering the external-space contribution at each threshold, the EN correction brings this sequence closer to the CI--MBPT limit than the variational energies alone and allows accurate energies to be reached with smaller variational spaces.

At $\epsilon=1.5\times10^{-3}E_{\mathrm h}$, CHCI--MBPT deviations from the CI--MBPT benchmark remain below 1\% for the optical excitation energies considered here, with maxima of 0.70\% for Si and 0.32\% for P.
This accuracy for optical excitation energies on the scale of $10^4\ \mathrm{cm}^{-1}$ motivates extending the calculations to larger CI spaces.
Adding higher virtual orbitals and further excitation classes is expected to introduce many weakly coupled configurations, offering the prospect of greater space compression while maintaining comparable excitation-energy accuracy.
\section{Conclusions}
\label{sec:conclusions}

We have developed CHCI--MBPT for multivalent atoms, combining a threshold on coherent couplings with whole-configuration selection in a CSF basis of definite $J^\pi$.
The Al calculations establish a systematic relation between the selection threshold, excitation-energy accuracy, and variational-space size.
The smaller compression factors for Si and P reflect both their stronger configuration mixing and the restriction of their parent spaces to single and double excitations from two reference configurations.
The benefit of selecting a separate space for each state depends on how much of the correlation space is shared by the target levels.

The external-space EN correction is central to the accuracy of these compact expansions.
It reduces the maximum and root-mean-square excitation-energy deviations for every tested atom, threshold, and selection mode, recovering correlation contributions without enlarging the space that is diagonalized.
The looser-threshold results demonstrate that this recovery remains effective when a larger fraction of the parent space is excluded from the variational calculation.
This combination of variational selection and perturbative recovery provides a basis for extending the calculations to larger orbital sets and excitation spaces, where weakly coupled configurations may offer further opportunities for compression.
\begin{acknowledgments}
We acknowledge support from the Strategic Priority Research Program of the Chinese Academy of Sciences (Grant No. XDB0920403), the National Natural Science Foundation of China (Grant No. 12393821), and the Innovation Fund for Scientific and Technological Personnel of Hainan Province (Grant No. 225RC636).
\end{acknowledgments}

\section*{Data Availability}
The data that support the findings of this article are not publicly available.
The data are available from the authors upon reasonable request.
\appendix
\section{Threshold dependence of Al~I and Al~II total energies}
\label{app:al-total-energies}
\begin{figure}[!htbp]
\centering
\includegraphics[width=\columnwidth]{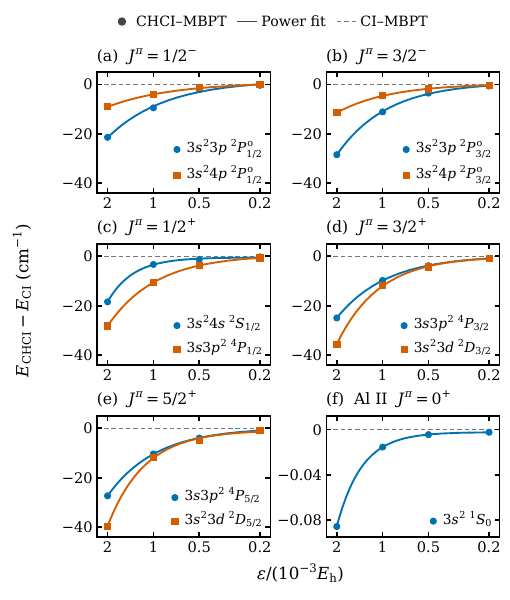}
\caption{Total-energy deviations of CHCI--MBPT from CI--MBPT for (a)--(e) Al I and (f) the Al II ground state.
Symbols are calculations, solid curves are fits to Eq.~\eqref{eq:al-energy-fit} relative to CI--MBPT, and dashed lines mark zero deviation.
The threshold axis is logarithmic and the energy axis is linear; Al II uses a smaller vertical scale.}
\label{fig:al-total-energy}
\end{figure}
\begin{table}[!htbp]
\caption{Parameters of the total-energy fits in Eq.~\eqref{eq:al-energy-fit}. The final column gives the root-mean-square fit residual.}
\label{tab:al-total-fit}
\small
\setlength{\tabcolsep}{2.5pt}
\begin{ruledtabular}
\begin{tabular}{lrrrr}
State & $A$ & $p$ & $E_{\mathrm{CI}}-E_0$ & Fit RMS \\
 & $(10^{-5}E_{\mathrm h})$ & & \multicolumn{2}{c}{$(\mathrm{cm}^{-1})$} \\ \hline
\multicolumn{5}{l}{Al I} \\
$3s^2 3p\;{}^{2}P^{\mathrm{o}}_{1/2}$ & $-4.86$ & $1.13$ & $-1.79$ & $0.61$ \\
$3s^2 4p\;{}^{2}P^{\mathrm{o}}_{1/2}$ & $-2.23$ & $1.03$ & $-0.86$ & $0.13$ \\
$3s^2 3p\;{}^{2}P^{\mathrm{o}}_{3/2}$ & $-5.41$ & $1.31$ & $-0.94$ & $0.21$ \\
$3s^2 4p\;{}^{2}P^{\mathrm{o}}_{3/2}$ & $-2.27$ & $1.25$ & $-0.34$ & $0.04$ \\
$3s^2 4s\;{}^{2}S_{1/2}$ & $-1.34$ & $2.61$ & $0.44$ & $0.22$ \\
$3s3p^2\;{}^{4}P_{1/2}$ & $-5.13$ & $1.36$ & $-0.67$ & $0.08$ \\
$3s3p^2\;{}^{4}P_{3/2}$ & $-4.62$ & $1.32$ & $-0.35$ & $0.06$ \\
$3s^2 3d\;{}^{2}D_{3/2}$ & $-5.53$ & $1.55$ & $-0.11$ & $0.13$ \\
$3s3p^2\;{}^{4}P_{5/2}$ & $-4.89$ & $1.36$ & $-0.29$ & $0.05$ \\
$3s^2 3d\;{}^{2}D_{5/2}$ & $-5.14$ & $1.78$ & $0.74$ & $0.47$ \\
\hline
\multicolumn{5}{l}{Al II} \\
$3s^2\;{}^1S_0$ & $-6.03\times10^{-3}$ & $2.66$ & $2.33\times10^{-3}$ & $6.62\times10^{-5}$ \\
\end{tabular}
\end{ruledtabular}
\end{table}
\begin{table}[!htbp]
\caption{First ionization energy of Al.}
\label{tab:al-ionization}
\begin{ruledtabular}
\begin{tabular}{lr}
Method & Ionization energy (eV) \\ \hline
Extrapolated & 5.9827 \\
$\epsilon=2\times10^{-4}\,E_{\mathrm h}$ & 5.9829 \\
NIST~\cite{NISTASD} & 5.9858 \\
\end{tabular}
\end{ruledtabular}
\end{table}

The total energies provide a direct way to examine threshold extrapolation.
Figure~\ref{fig:al-total-energy} shows their signed deviations from CI--MBPT for the Al I levels and the Al II ground state.
The latter calculation uses the same frozen $[\mathrm{Ne}]$ core and orbital spaces, with two active valence electrons.
For each state, we fit the four calculated energies to
\begin{equation}
E(\epsilon)=E_0+A\left(\frac{\epsilon}{\epsilon_{\mathrm{ref}}}\right)^p,
\label{eq:al-energy-fit}
\end{equation}
where $\epsilon_{\mathrm{ref}}=10^{-3}E_{\mathrm h}$, $E_0$ is the zero-threshold limit, $A$ is an energy coefficient, and $p>0$ is the fitted exponent.
All three parameters are determined by least squares in total energy.

The fitted parameters and residuals are listed in Table~\ref{tab:al-total-fit}.
The present four-point power-law fits do not improve the Al I excitation energies.
The direct results at the tightest threshold, $\epsilon=2\times10^{-4}E_{\mathrm h}$, are more accurate for all nine excited levels, with a mean absolute deviation of $0.32\ \mathrm{cm}^{-1}$ compared with $1.53\ \mathrm{cm}^{-1}$ after extrapolation.
This may reflect both the limited number of points available to constrain the fits and the empirical power-law form, for which a rigorous theoretical justification of the threshold dependence has not been established.
Nevertheless, the extrapolation errors remain small, with a maximum excitation-energy deviation of $2.53\ \mathrm{cm}^{-1}$, and are acceptable for the optical excitation energies considered here.
Threshold extrapolation therefore remains a promising route when direct CHCI--MBPT calculations at sufficiently tight thresholds become too costly.
For the simpler two-valence-electron Al II system, the extrapolated energy agrees with CI--MBPT to $2.33\times10^{-3}\ \mathrm{cm}^{-1}$.
Table~\ref{tab:al-ionization} gives the ionization energy obtained from the difference between the Al II and Al I ground-state energies, using either the fitted limits or the results at $\epsilon=2\times10^{-4}E_{\mathrm h}$, together with the NIST reference value.
\bibliography{main}
\end{document}